\documentclass{aa}  
\usepackage{graphicx}
\usepackage{txfonts}
\usepackage[T1]{fontenc}
\usepackage[utf8]{inputenc}
\usepackage{todonotes}
\usepackage{color}
\usepackage{xcolor}
\usepackage{hyperref}
\hypersetup{colorlinks=true,citecolor=blue}
\usepackage{tabularx}
\usepackage[normalem]{ulem} 
\usepackage{xcolor}

\begin{document}

   \title{Characterisation of the continuum and kinematical properties of nearby NLS1}

   \author{Gabriel A. Oio\inst{1}
          \and Luis R. Vega\inst{1,2}
          \and Eduardo O. Schmidt\inst{1,2}
          \and Diego Ferreiro\inst{1,2}
           }
   \institute{Instituto de Astronom\'ia Te\'orica y Experimental (IATE), CONICET - UNC, Laprida 854,
   X5000BGR,C\'ordoba, Argentina\\
             \and
             Observatorio Astron\'omico, Universidad Nacional de C\'ordoba, Laprida 854, X5000BGR, C\'ordoba, 
             Argentina.\\
              }
   \authorrunning{Oio et al.}
 
  \abstract
   {}
   {In order to study the slope and strength of the non-stellar continuum, we analysed a sample from nearby Narrow Line Seyfert 1 (NLS1). Also, we re-examined the location of NLS1 galaxies on the $M_{BH}$ - $\sigma_{\star}$ relation, using the stellar velocity dispersion and the [OIII]$\lambda$5007 emission line as a surrogate of the former.
   }
   {We studied spectra of a sample of 131 NLS1 galaxies taken from the Sloan Digital Sky Survey (SDSS) DR7. We approached determining the non-stellar continuum by employing the spectral synthesis technique, which uses the code {\sc starlight}, and by adopting a power-law base to model the non-stellar continuum. Composite spectra of NLS1 galaxies were also obtained based on the sample. In addition, we obtained the stellar velocity dispersion from the code and by measuring Calcium II Triplet absorption lines and [OIII] emission lines. From Gaussian decomposition of the H$\beta$ profile we calculated the black hole mass.}
   {We obtained a median slope of $\beta$ = $-$1.6 with a median fraction of contribution of the non-stellar continuum to the total flux of 0.64. We determined black hole masses in the range of log(M$_{BH}$/M$_{\odot}$) = 5.6 $-$ 7.5, which is in agreement with previous works. We found a correlation between the luminosity of the broad component of H$\beta$ and black hole mass with the fraction of a power-law component. Finally, according to our results, NLS1 galaxies in our sample are located mostly underneath the $M_{BH}$ - $\sigma_{\star}$ relation, both considering the stellar velocity dispersion ($\sigma_{\star}$) and the core component of [OIII]$\lambda$5007.}
   {}


   \maketitle
%

\section{Introduction}
\label{Intro}
A characteristic feature of active galactic nuclei (AGN) is the presence of energy outputs, which are not related to ordinary stellar processes \citep{1969Natur.223..690L}. The presence of a non-stellar component is evident in AGN spectra, and it is frequently expressed in the form of a power-law $f_{\lambda} \propto \lambda^{\beta}$ or $f_{\nu} \propto \nu^{\alpha}$ with the spectral index $\alpha=-\beta-2$. Though a rough approximation, this expression is very useful in describing the continua of AGN. For low redshift AGN, a value of $\beta=-1.3$ was adopted in the 1980s \citep[e.g.][]{1980ApJ...235..361R,1988MNRAS.233..801O,1989ApJS...69..703S}, and later a harder slope with $\beta = -1.7$ was claimed \citep{1987ApJS...63..615N,1991ApJ...373..465F}. Through accurate Near-infrared (NIR) optical photometry for a sample of bright Quasi-Stellar Objects (QSOs) at $z$ $\sim$ 2, \cite{1996PASA...13..212F} determined a median slope of $\beta = - 1.65$, which is consistent with the free-free emission models of \cite{1993ApJ...412..513B}. For high-redshift quasars, the disparity in the determination of $\beta$ is critical as it introduces uncertainties in their measured evolution, which was pointed out by \cite{1996PASA...13..212F}. For instance, an adopted value of $\beta = -1.7$ instead of $-$1.5 implies that, due to difference in $k-$corrections, the luminosity function
should be incremented by a factor of two at $z\sim$ 2, thus altering the inferred evolution of QSO number densities \citep{1992ApJ...396..411G,2000A&A...353..861W}.

Most of the aforementioned determinations of the non-stellar continuum were made through photometry for quasars and in the UV range. In addition, spectral measurements were done in order to characterise the continua of AGNs. By creating composite quasar spectra, \cite{2001AJ....122..549V}, obtained slopes of $\beta = -1.54$ and  $-$0.42 for UV and optical ranges, respectively. \cite{2007ApJ...668..682D} obtained values of $\beta$ mostly in the range between approximately $-$1.5 to $-$2, which were restricted to the UV range. Despite these important efforts, lately, there have been few works that focus on the determination of the spectral index in the optical for low-redshift AGNs. Also, the classical value of $\beta = -1.5$ is generally still widely adopted  \citep{2013MNRAS.433.1161A,2009RAA.....9.1078W,2008ApJ...685..801Y}. An exception would be the recent spectral analysis of \cite{2017MNRAS.472.4051C}, who obtained spectral slopes for Sloan Digital Sky Survey (SDSS) spectra; however, they assume a fixed value of $\beta = -1.7$ for low-redshift (z $\le$ 0.6) AGNs due to the difficulty of separating stellar and non-stellar components in AGNs closer than z $\sim$ 0.7. The presence of the stellar contribution in the measured flux of nearby AGNs is an important caveat in determining the spectral index. When performing spectroscopy, this contribution strongly depends on which fraction of the galaxy is observed, which in turns depends on the width of the fibre used (or slit and corresponding spectral extraction) and, therefore, on the distance of the observed galaxy. For example, an extraction of 1 arcsec wide in a galaxy at 100 Mpc corresponds to a projected distance of about 500 pc, whilst for further galaxies we would be mapping a region of several kpc. Thus, in order to determine the non-stellar continuum it is necessary to properly discount the stellar contribution matching those regions. This effect may only be minimum for luminous bright QSOs, for which the brightness is so high that the emission from their host galaxy could be neglected \citep{1996PASA...13..212F, 2007ApJ...668..682D, Donley2010, Donoso2018}; however, care must be taken when assuming this for lower luminosity
AGNs.

Narrow Line Seyfert 1 galaxies \cite[NLS1;][]{1985ApJ...297..166O,1989ApJ...342..224G} are an interesting subclass of AGN with relatively narrow widths of permitted optical emission lines. These galaxies are defined from their optical spectral characteristics as follows: full width at half maximum (FWHM) of H$\beta$ $\le$ 2000 km s$^{-1}$, ratio [OIII]$\lambda$5007/H$\beta$ $\le$ 3, and strong FeII multiplets (see \citealt{2008RMxAC..32...86K} for a review).
In this paper we focus on the determination of the spectral index of the non-stellar continuum in NLS1, as well as the stellar and non stellar contributions to the total emission observed in the optical range.

For NLS1s, the less massive black holes (BH) were obtained and then associated to the highest accretion rates, suggesting that they might be in the early stage of AGN evolution \citep{Grupe1999,2000MNRAS.314L..17M,2000NewAR..44..469M,2007ApJ...660.1072W}.
The well-known scaling relation between black hole mass and host galaxy stellar velocity dispersion ($M_{BH}$ - $\sigma_{\star}$) may indicate a co-evolution of black hole and galaxy \citep{2000ApJ...539L...9F,2013ARA&A..51..511K}. 
For NLS1s, this behaviour has been explored by many authors who have found diverse results. While normal galaxies and AGN have similar slopes in this relation, NLS1s seem to deviate from it in the sense that they should have smaller black hole masses for a given stellar velocity dispersion \citep{2001NewA....6..321M,2004MNRAS.347..607B,2004ApJ...606L..41G,2005ApJ...633..688M,2005A&A...432..463M,2006ApJS..166..128Z,Schmidt2016,2017ApJS..229...39R}. On the contrary, other authors claimed that NLS1s are `on' the $M_{BH}$ - $\sigma_{\star}$ relation \citep{2001A&A...377...52W,2005MNRAS.356..789B,2007ApJ...667L..33K,2016MNRAS.462.1256C}. Sometimes this could be due to the difficulty of directly measuring the stellar velocity dispersion in AGNs. To circumvent this, some emission lines, such as [OIII]$\lambda$5007 or [SIII]$\lambda$9069, could be used as a surrogate of $\sigma_{\star}$ \citep{1996ApJ...465...96N,Shields2003,2009MNRAS.393..846V}.

Here we re-examine the $M_{BH}$ - $\sigma_{\star}$ relation for NLS1s using $\sigma_{\star}$ that are directly measured from the Ca II Triplet absorption line and the one obtained from stellar synthesis templates. Furthermore, we tested the use of the `core' of [OIII]$\lambda$5007 emission line, after removal of asymmetries, as a replacement for the stellar velocity dispersion.

In this paper, our main aims are firstly to measure the spectral index $\beta$ in the optical region and the non-stellar contribution to the observed spectra for a sample of 131 nearby NLS1s taken from the Seventh Data Release of the Sloan Digital Sky Survey (SDSS - DR 7) \citep{2000AJ....120.1579Y, 2009ApJS..182..543A}, and secondly to study some properties inferred from the emission lines with the objective of revisiting the locus of NLS1 galaxies in the $M_{BH}$ - $\sigma_{\star}$ relation found for normal galaxies \cite[e.g.][]{2000ApJ...539L...9F, 2000ApJ...539L..13G,2002ApJ...574..740T}.
The sample and data treatment is described in section \ref{Data}.  General results are presented in section \ref{Results} and a discussion appears in
section \ref{Discussion}. Throughout this paper we assume H$_{\rm o}$=71 km s$^{-1}$ Mpc$^{-1}$, $\rm \Omega_{m}$=0.3 and $\rm \Omega_{\Lambda}$=0.7.


\section{Data analysis}
\label{Data}

\subsection{Sample selection}
\label{Sample}

\begin{figure}
\begin{center}
\includegraphics[width=\columnwidth]{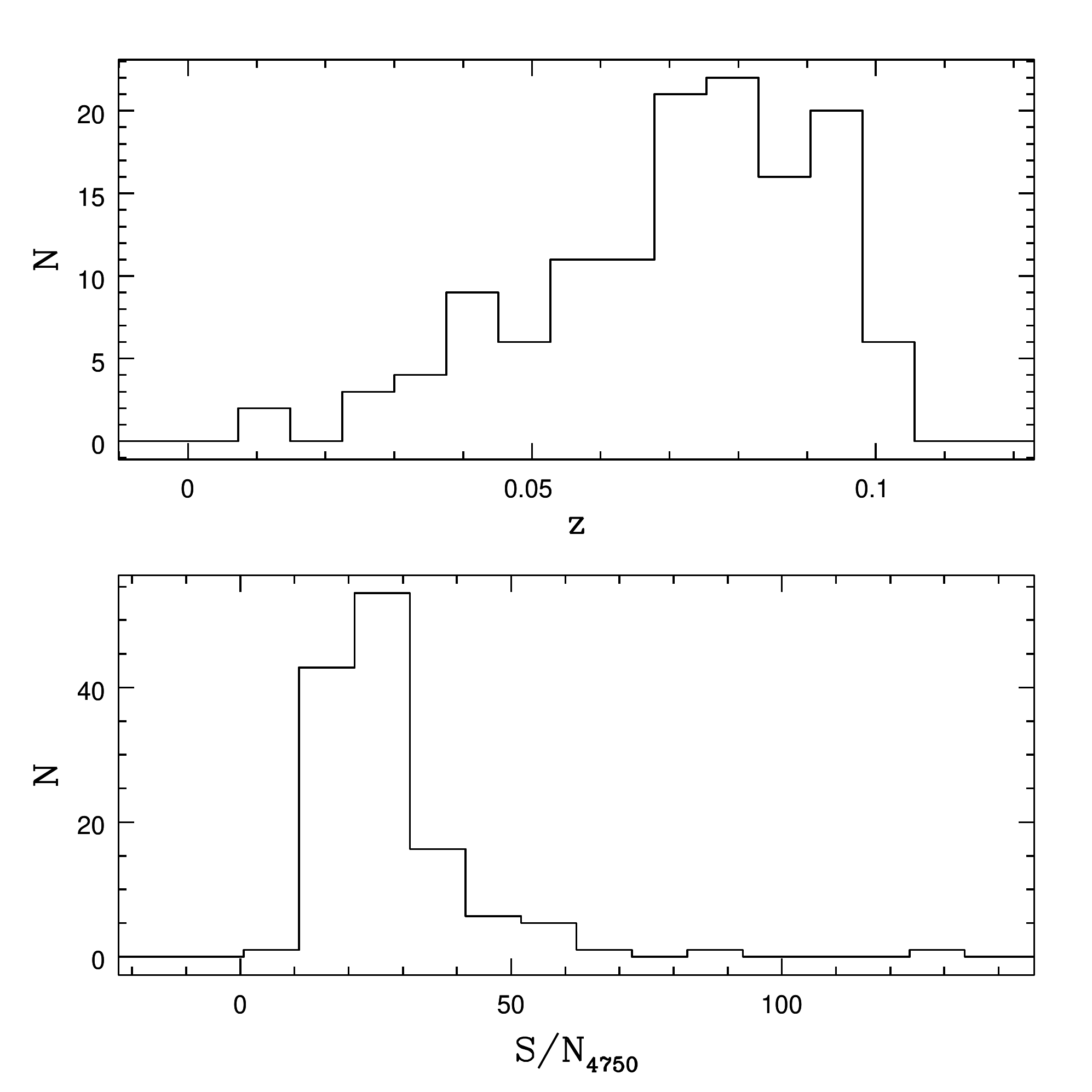}
\end{center}
\caption{Distributions of redshift (top panel) and S/N of continuum measured at rest-frame 4750\AA \ (bottom panel).}
\label{fig:Hist distribution}
\end{figure}

Our sample consists of 131 NLS1 at z $\le$ 0.1 selected from the list of \cite{2006ApJS..166..128Z}, which is also available in the catalogue of \cite{2010A&A...518A..10V}. By modelling the emission lines and continua of active objects labelled as `QSOs' or `galaxies' in the Third Data Release of the Sloan Digital Sky Survey (SDSS - DR3), \citep{2005AJ....129.1755A}, \cite{2006ApJS..166..128Z} have performed an exhaustive search and found 2011 NLS1 up to z$\sim$0.8. This limit corresponds to the detection of H$\beta$ in SDSS spectral coverage ($\sim$ 3800 \AA\ to 9200 \AA). We chose 131 NLS1 from the whole list corresponding to objects nearer than z=0.1 (Fig. \ref{fig:Hist distribution}, top panel).
The spectroscopic observations were taken from SDSS DR7, which is complete to a Petrosian magnitude \citep{1976ApJ...209L...1P} of 17.77 mag \citep{2002AJ....124.1810S}. The spectra have a resolution $\sim$ 2000, and for our sample we measured a mean Signal to Noise (S/N) $\sim 25$ obtained in the window $\lambda$4730 $-$ $\lambda$4780 (Fig. \ref{fig:Hist distribution}, bottom panel).
In the redshift range of our sample, the SDSS fibre aperture of 3 arcsec correspond to 0.79 $-$ 5.45 kpc of the projected distance, so we are dealing with integrated spectra.

Host galaxy contamination in SDSS spectra can be noticeable even in the case of some quasars \citep{2001AJ....122..549V}.
After visual inspection of the sample, we found that in most cases absorption lines are easily recognised. For instance the CaK $\lambda$3933 line is observed in 96 galaxies, while the Na-D lines $\lambda \lambda$5889.9, 5895.9 can be seen in 62 cases. The Ca triplet is mostly restricted due to the redshift of the galaxy. Ca II $\lambda$8498 and CaII$\lambda$8542 were detected in 53 objects and 59 spectra, respectively, while CaII$\lambda$8662
was found in 23 out of 131 objects. Careful removal of the stellar contribution is essential in making proper determinations of the non-stellar continuum and reliable measurements of the recombination emission lines. Furthermore, this 'starlight contamination' also provides valuable information about the host galaxies, which is of interest itself. For this, in addition to other purposes, we have developed a technique to properly model the stellar and non-stellar components, which is explained in the next subsection. 

\subsection{Continuum modelling}
\label{Modeling}

All spectra were corrected for galactic extinction values available in
NED (NASA/IPAC Extragalactic Database) and were calculated according to \cite{1998ApJ...500..525S}. After extinction correction, we shifted the spectra to rest-frame using redshift values given by SDSS pipelines and re-sampled the spectra in steps of $\Delta\lambda$ = 1 \AA{}.
We modelled the continuum of the galaxies using the spectral synthesis code {\sc starlight} \citep{2005MNRAS.358..363C, 2006MNRAS.370..721M}. This code was widely used to model the stellar populations of SDSS galaxies \citep{2005MNRAS.356..270C} and different samples of AGN \citep[e.g,][]{2004MNRAS.355..273C,2007ASPC..373..675B,2009MNRAS.393..846V,2011MNRAS.411.1127L,2013ApJ...763...36B}.
Basically, {\sc starlight} models the observed spectrum $O_{\lambda}$
to obtain the model $M_{\lambda}$ by a linear combination of simple
stellar populations (SSPs), as

\begin{equation}
M_\lambda(x,M_{\lambda_0},A_V,v_\star,\sigma_\star) = M_{\lambda_0}
\left[\sum_{j=1}^{N} x_j b_{j,\lambda} r_\lambda \right] \otimes
G(v_\star,\sigma_\star)
\label{eq:Stl}
\end{equation}

\noindent where $b_{j,\lambda} \equiv L_j(\lambda) / L_j(\lambda_0)$ is the $j$th SSP normalised at $\lambda_0$; $x_j$ are the components of the `population vector'; $M_{\lambda_0}$ is the synthetic flux at $\lambda_0$; $r_\lambda = 10^{-0.4(A_\lambda-A_{\lambda_0})}$ is the reddening term; $A_\lambda$ is the internal extinction at $\lambda$ for each object, which is modelled as a dust screen in the line of sight and parametrised in terms of $A_V$ (extinction in $V$-band), adopting $R_V$ = 3.1
\citep{1989ApJ...345..245C}; $N$ is the total number of components in the spectral base; and $G(v_\star,\sigma_\star)$ is the line of sight stellar velocity distribution modelled as a Gaussian centred at $v_\star$ with a velocity dispersion $\sigma_\star$. The spectra of the base ($b_{j,\lambda}$) are convolved ($\otimes$) in order to take into account the absorption line broadening. The best fit is obtained by means of minimisation of $\chi^{2}$,

\begin{equation}
\chi^2 \equiv \sum_{\lambda_i}^{\lambda_f} [O_{\lambda} - M_{\lambda}]^2.\omega_{\lambda}^2
\label{chi2}
\end{equation}

\noindent where $O_{\lambda}$ is the observed spectrum, $\omega_{\lambda}$ is the weight defined as the inverse of the noise in $O_{\lambda,}$ and $\lambda_{i,f}$ are the initial and final wavelengths. If one does not want to model spectral regions with {\sc starlight} (e.g. emission lines), they are masked out by choosing $\omega_{\lambda}=0$. Besides the high contribution due to the non-stellar continuum and the emission lines, NLS1 spectra often show some absorption lines, mainly Ca K, Ca H, and CaII triplet. These absorption features are given positive weights in the mask (i.e. $\omega_{\lambda}>0$) to be sure that the stellar contribution is properly fitted. See \cite{2004MNRAS.355..273C,2005MNRAS.358..363C, 2006MNRAS.370..721M} for more details.

When performing this type of analysis, it is important to define the base components. To account for the stellar contributions, 80 SSPs of \cite{2003MNRAS.344.1000B}
were adopted, corresponding to 20 ages and four metallicities. The ages are the following: $t_\star$ = 0.00316, 0.00501, 0.00661, 0.00871, 0.01, 0.01445, 0.02512, 0.04, 0.055, 0.10152, 0.1609, 0.28612, 0.5088, 0.90479, 1.27805, 1.434, 2.5, 4.25, 6.25, and 7.5 $\times 10^9$ years, while the metallicities are $Z$ = 0.2, 0.4, 1, and 2.5 $Z_\odot$. Besides these stellar components, a power-law (PL) component should be added to the spectral base since we are dealing with active nuclei. This PL component is usually represented as $F_{\nu} \sim \nu^{\alpha}$ or equivalently $F_{\lambda} \sim \lambda^{\beta}$, with $\alpha=-\beta-2$. Our contribution to the modelling of the non-stellar continuum of NLS1s is the inclusion of a base in the form $F_{\lambda}=10^{20}\times(\lambda/4020)^{\beta}$, corresponding to six spectral indexes $\beta$ ranging from -3 to -0.5 with steps of 0.5. As discussed in \cite{2004MNRAS.355..273C}, noise in the spectra dilutes the differences in the base components, which are similar due to intrinsic degeneracies. A way to avoid this is by grouping the $x_j$ of similar spectral components, which would return a more reliable result.
Thus, the fraction of the observed flux due to the power-law component can be parametrised as:
\begin{equation}
 F_{PL} \sim \sum_{i=1}^{6}X_{PL_i}\lambda^{\beta_i}
 \label{eq:FPL}
\end{equation}
from which we can derive a single power-law with a mean spectral index $<\beta>$ for each galaxy. We do not seek to assign any physical meaning to this PL base component since it is merely included to add a free parameter to the fit. 

\begin{figure*}
 \begin{minipage}{\linewidth}
  \begin{center}
\includegraphics[trim = 0mm 0mm 0mm 1mm, clip,width=\textwidth]{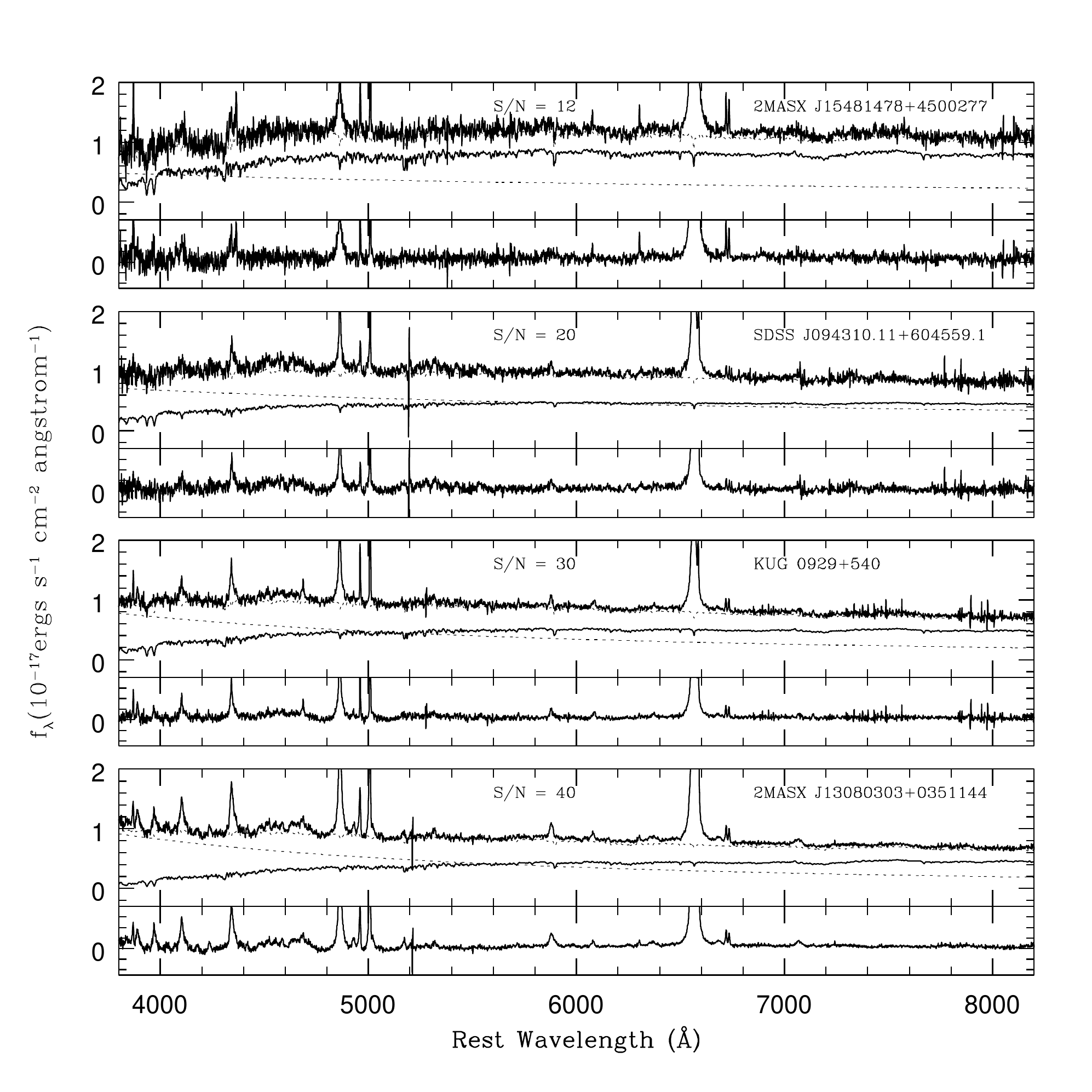}
\caption[Fits of AGNs 1]{Representative examples of spectral fitting. In each panel, we plot the observed spectrum (thick solid line), the modelled spectrum (dotted line), the host galaxy spectrum (thin solid line), the power-law contribution (dotted line), and the residual spectra (lower panel). The observed spectra are normalised at the flux value in $\lambda$ = 4020. = 4020.}
\label{Fig:fits2}
 \end{center}
 \end{minipage}
\end{figure*}

This method allow us to identify the following three important parts of an AGN spectrum: the stellar populations, the non-stellar continuum, and the difference between the total flux and the sum of these two components that give us the residual spectra, that is, the ionised interstellar gas.
The search for the stellar and non-stellar contributions is made by computing equation \ref{chi2} for each galaxy after masking the emission lines. 
We show in Figure \ref{Fig:fits2} examples of spectral synthesis for selected galaxies. 
An estimate of the errors in the parameters acquired from {\sc starlight} were obtained using the jacknife technique. For this, we ran the code 85 times, taking away one component from the base in each run. The adopted error was the dispersion (at 1 $\sigma$) in the distribution of the measurement of each parameter. From this we have a typical uncertainty of $\pm$ 0.2 for $\beta$ and of $\sim$ 5\% for F$_{PL}$. For those objects with a contribution of F$_{PL}$ that is less than $\sim$ 50 \%, the uncertainties in these parameters are larger.

\begin{figure*}
\begin{minipage}{\linewidth}
\begin{center}
\includegraphics[trim = 0mm 5mm 0mm 100mm, clip,width=\textwidth]{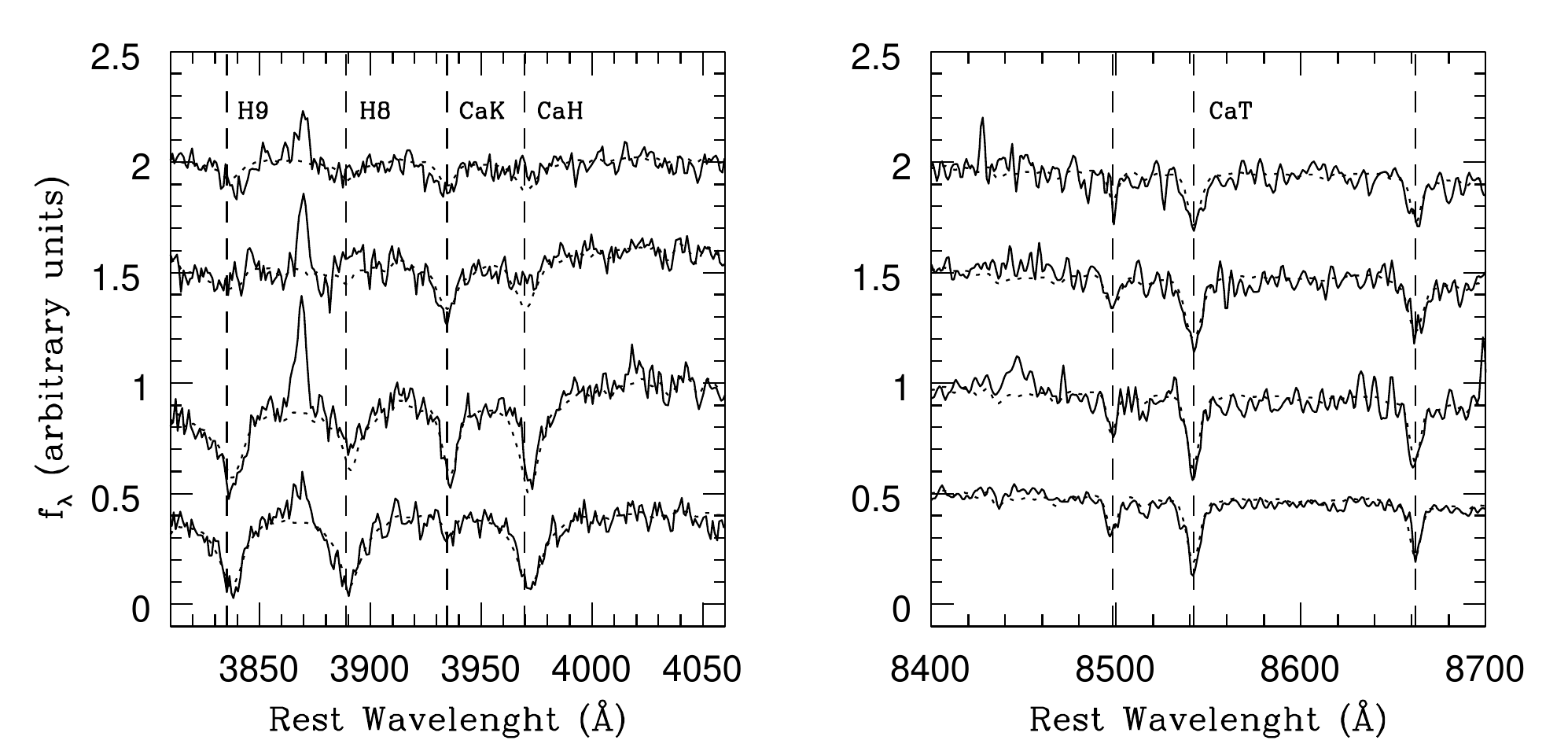}
\caption{Example of NLS1 galaxies with noticeable absorption lines. The observed continuum and the synthetic spectra obtained with STARLIGHT correspond to the solid line and the dotted line, respectively.}
\label{fig:verabs}
\end{center}
\end{minipage}
\end{figure*}

\subsection{Absorption lines}
As previously mentioned, a significant portion of our sample presented absorption lines in their observed spectra. We chose to measure the ionised calcium K and calcium triplet lines because they are the stellar features with the highest S/N and the least amount of contamination due to emission lines or non-stellar continuum.
In the case of Ca II triplet (CaT), we used the penalised pixel-fitting method (pPXF) \cite{2017MNRAS.466..798C} to extract the stellar kinematics. We fitted the kinematics of CaT, using the MILES stellar library \citep{2016MNRAS.463.3409V}, in the range 8350$-$9020\AA{} with a resolution of 1.5\AA \ \. (FWHM). 
On the other hand, the Ca K absorption line was measured using the FITPROFS task included in IRAF\footnote{IRAF: the Image Reduction and Analysis Facility is distributed by the National Optical Astronomy Observatories, which is operated by the Association of Universities for Research in Astronomy, Inc. (AURA) under cooperative agreement with the National Science Foundation (NSF) \cite{1993ASPC...52..173T}.}, assuming that a single Gaussian function was sufficient to fit the profile. 
All FWHM were corrected for instrumental broadening (FWHM$_{inst}$) considering the resolution given by Sloan of FWHM = 2.7 \AA{} at $\lambda$ = 4861. In Fig. \ref{fig:verabs} we show examples of spectra with prominent absorption lines, and in Fig. \ref{fig:Hist distribution} (bottom panel) we show the equivalent width (EW) distribution for the CaK and CaII$\lambda$8542 lines. The EW of CaK has a mean value of 2.8 $\pm$ 1.9 and a median of 2.6 with an IQR of 2.5, while the EW of CaII$\lambda$8542 has a mean value of 2.2 $\pm$ 0.9 and a median of 2.2 with an IQR of 1.3.

\begin{figure}
\begin{center}
{\setlength{\fboxsep}{0pt}
 \setlength{\fboxrule}{0pt}
 \fbox{\includegraphics[trim = 0mm 5mm 0mm 100mm, clip,width=\columnwidth]{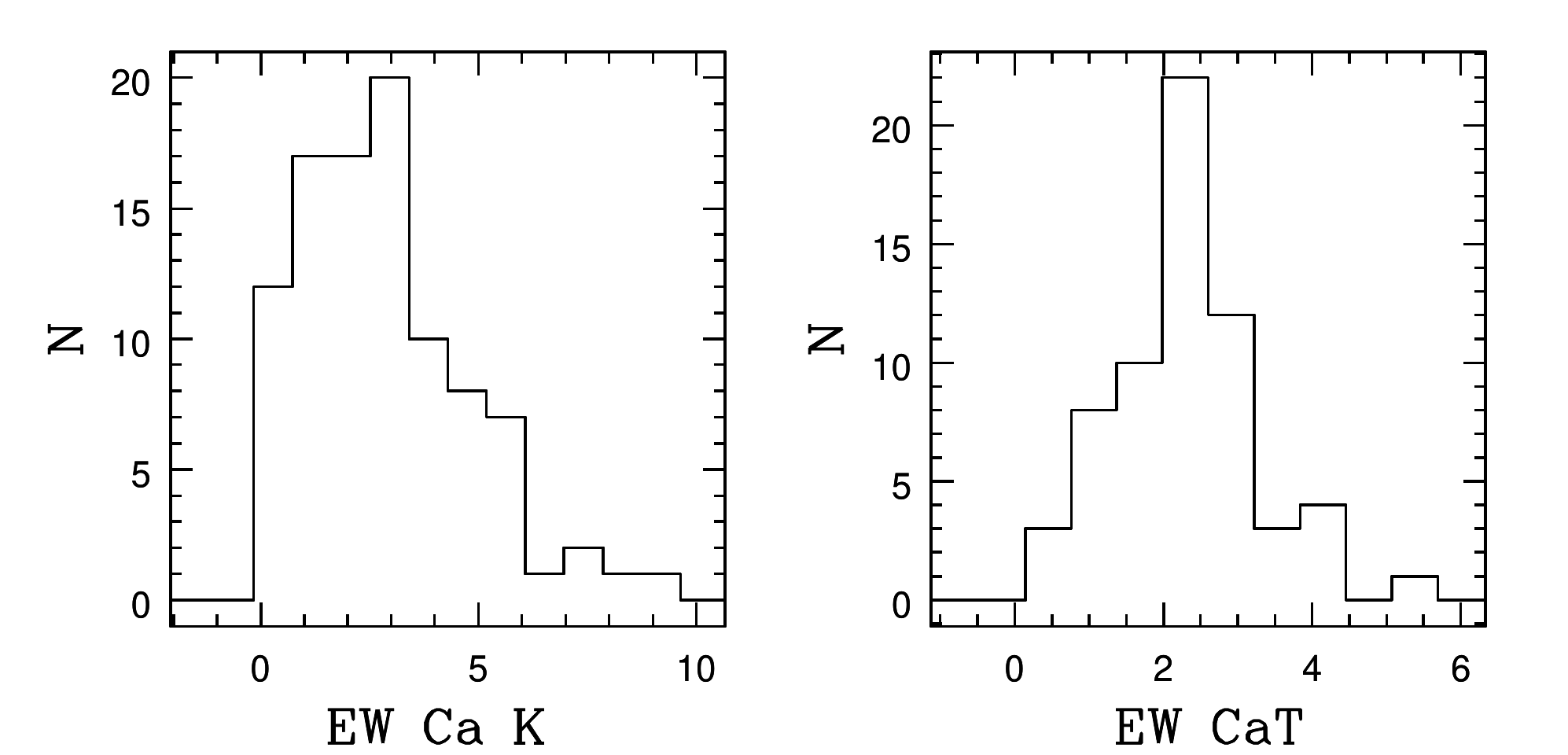}}
}
\caption{Distributions of equivalent width of CaK line (left) and CaII$\lambda$8542 line (right) for our sample.}
\label{fig:CaK distribution}
\end{center}
\end{figure}

\subsection{Iron emission lines}
\label{Emission Lines}

\begin{figure}
\begin{center}
\includegraphics[width=\columnwidth]{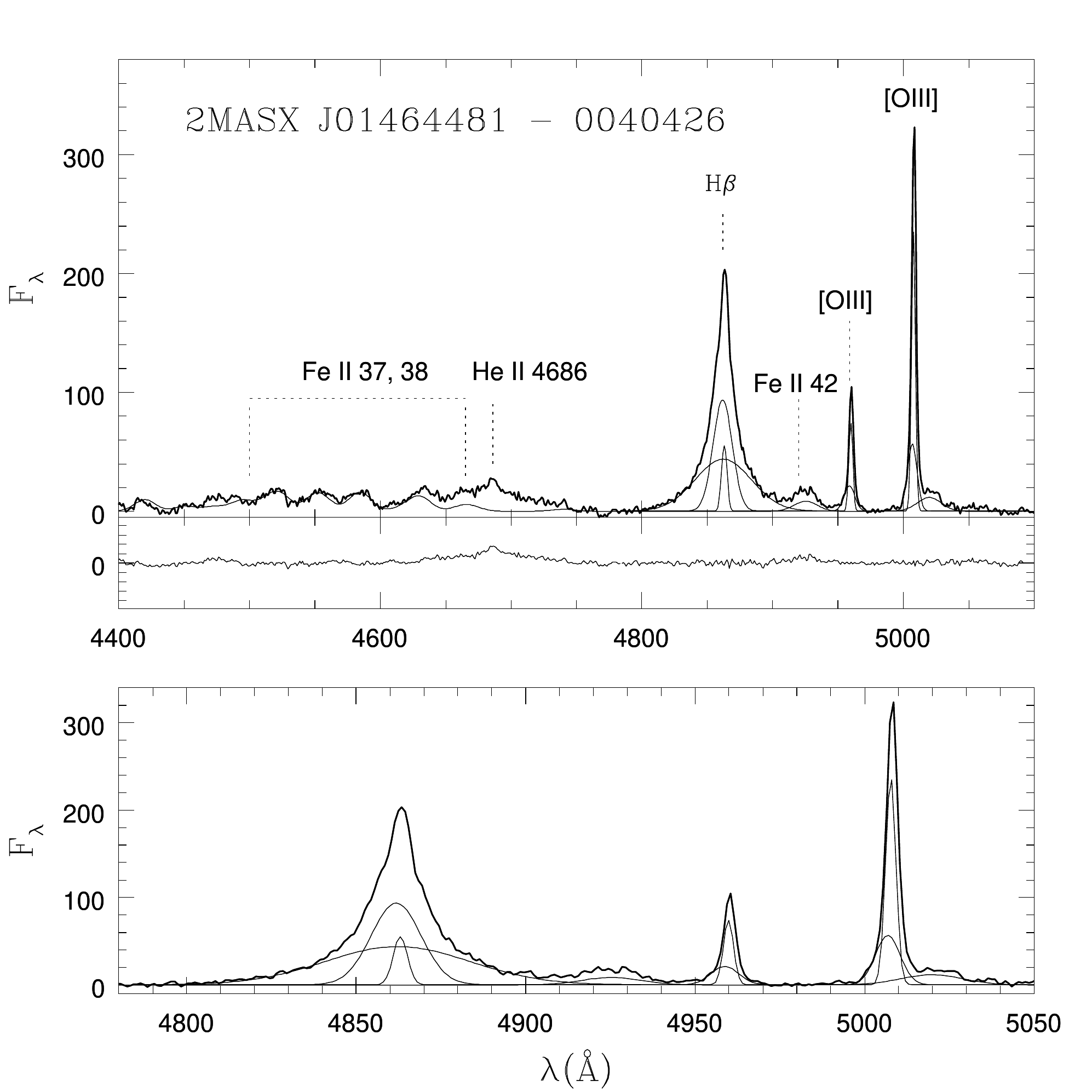}
\end{center}
\caption{Top panel shows Gaussian decomposition and iron fit for galaxy 2MASX J01464481-0040426. The observed spectrum flux is in arbitrary units (thick line); FeII template and Gaussian components obtained with LINER are in thin lines. The residual spectra are plotted on the lower sub-panel for clarity. The bottom panel shows a zoomed in view of the H$\beta$+[OIII] observed spectrum and its fit.}
\label{fig:GaussFit}
\end{figure}

The primary challenge in measuring accurate fluxes of emission lines (see sec. \ref{Gauss}) is to properly subtract any kind of contamination.
It is interesting to note that NLS1s usually show strong optical FeII emission lines \citep[e.g.][]{2004A&A...417..515V,2006ApJS..166..128Z,2016MNRAS.462.1256C}. Furthermore, some authors even suggest adopting the flux ratio of FeII instead of H$\beta$ because it is better at differentiating NLS1s from BLS1s \citep{2001A&A...372..730V}. We focused part of our analysis on the spectral emission lines on the H$\beta$ region, between 4000 and 5500\AA{}, where strong FeII multiplets are often present. We subtracted the iron emission lines using the online software developed by \cite{2010ApJS..189...15K} and \cite{2012ApJS..202...10S}\footnote{\url{http://servo.aob.rs/FeII_AGN/}}. This software provides a best-fit model that reproduces the iron multiplets in the H$\beta$ region for each object in function of the gas temperature, Doppler broadening, and shift of the FeII lines. The software performs a $\chi^2$ minimisation routine to obtain the best fit. To fit each line, it assumes that it can be represented by a Gaussian, described by width, shift, and intensity. Given that these lines are likely to originate in the same region with the same kinematical properties, values of shift and width are assumed to be the same for all Fe II lines in the case of one AGN. See Fig. \ref{fig:GaussFit} for  an example.

\subsection{Line measurements}
\label{Gauss}

After subtraction of the stellar and non-stellar components and the Fe multiplets, we are left with the lines emitted by the ionised gas. Emission line profile fitting has been tackled in different ways by various authors. For low line-of-sight inclination systems \cite{2012MNRAS.426.3086G} concludes that turbulence dominates the planar Keplerian motion, resulting in Lorentzian profiles for lines formed at large BLR radii. From observational studies of NLS1 spectra,  some authors choose the
Lorentz profile in order to model the broad component of H$\beta$  \citep[e.g.][]{2006ApJS..166..128Z,2016MNRAS.462.1256C,2017ApJS..229...39R}. On the other hand, some authors suggest that a multiple Gaussian profile fitting yields a more statistically robust result than the Lorentzian component fitting method \citep{2005ApJ...623..700D,2008MNRAS.385...53M}.
In our analysis we assumed that the emission line profile of the broad component of H$\beta$ in NLS1 galaxies can be represented by a single or a combination of Gaussian profiles \cite[e.g.][]{2017ApJ...848...35S,Schmidt2016,2008MNRAS.385...53M,Schmidt2019}. For this purpose, we used the LINER routine \citep{1993Pogge..and..Owen}, which is a $\chi^2$ minimisation algorithm that can fit several Gaussians to a line profile. 
We fitted the [OIII] $\lambda$4959 and $\lambda$5007 lines with one Gaussian for the core component, and one or two additional Gaussian components for the asymmetric emission profiles, depending on the case \citep[e.g.][]{2016MNRAS.462.1256C}. In our sample, 28 galaxies showed no asymmetries and were fitted with only one component, 86 galaxies were fitted with one extra Gaussian for the asymmetric emission, while 15 galaxies needed the inclusion of two extra components. In a similar way, we also fitted H$\beta$, which is one of the strongest permitted lines in the optical range. We fitted the narrow component taking into account that it should have approximately the same FWHM as the core component of [OIII]$\lambda$5007 \citep{2018A&A...615A..13S} and one or two Gaussian components to fit the broad emission as mentioned.
The two galaxies, SDSS J144249.70+611137.8 and SDSS J103210.15+065205.3, show no emission in H$\beta$ and they were therefore excluded from the subsequent analysis. 
In order to asses the measurement uncertainties, we measured the emission lines at least 15 times in galaxies with different S/N. In the case of a galaxy with an S/N of $\sim$ 13, we found a relative error in the measured values of $\sim$20\% for the FWHM and of $\sim$4\% for the flux of the narrow component of H$\beta$ and 1\% and 3\% for the FWHM and flux errors of the broad component of H$\beta$. For the [OIII] lines, the uncertainties are of the order of $\sim$ 4\% and 9\% for the FWHM and flux measurements of the central component, and of 30\% and 20\% for the FWHM and flux of the asymmetric component. Considering galaxies with a typical S/N (approximately 26), the measurement errors are of 4\% for the FWHM and flux of the narrow component of H$\beta$, and they are of 10\% and 20\% for the FWHM and flux of the broad component of H$\beta$. In the case of the [OIII] line, the relative errors are of the order of 5\% and 10\% FWHM and flux respectively in the case of the core component, and 10\% and 2\% for the FWHM and flux of the asymmetric component. A typical fit is shown in Fig. \ref{fig:GaussFit}.


\section{Results}
\label{Results}
\subsection{Non-stellar content in AGNs}

\begin{figure*}
\begin{minipage}{\linewidth}
\begin{center}
\includegraphics[trim = 0mm 0mm 0mm 105mm, clip, width=\linewidth]{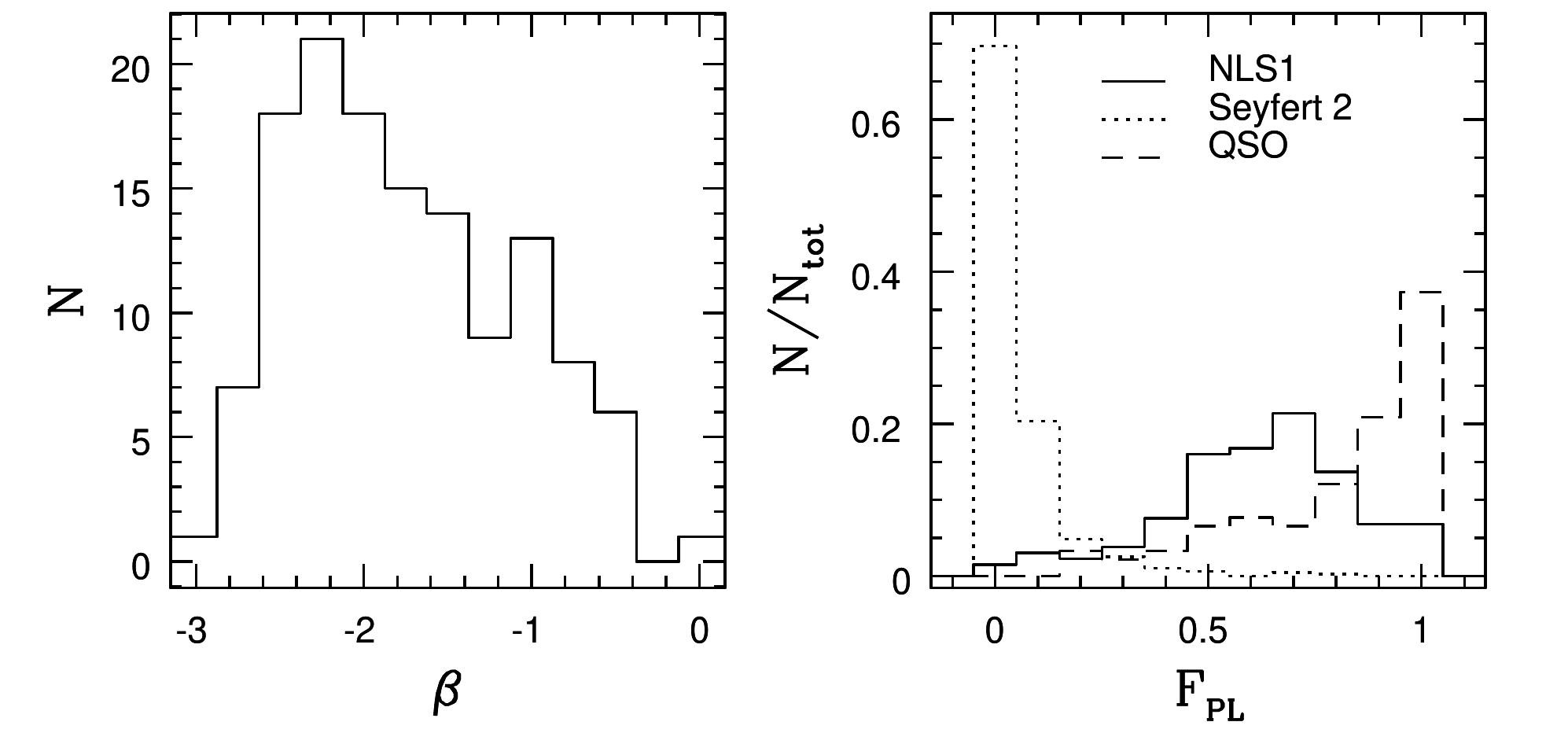}
\end{center}
\caption[Betas]{Distribution of power-law indexes obtained from {\sc starlight} for our sample of 131 NLS1 (left panel) and fraction of contribution of this component to the total spectra (right panel). Solid line represents NLS1, dotted line BLAGNs, and dashed line Seyfert 2 galaxies.}
\label{Fig:Betas}
\end{minipage}
\end{figure*}

Here we mainly focus on the non-stellar continuum of AGN, which were calculated using a spectral base of power$-$law, as described in section \ref{Modeling}. We stress that, unlike other works, we simultaneously obtained the non-stellar contributions, F$_{PL}$ and the optical slopes, for each AGN. This technique differs from what is accepted by other authors, for instance, who always assume the same spectral index \citep[e.g.][]{2006ApJS..166..128Z} or who consider that all of the observed continuum arise from the active nucleus, as may occur in quasars and Seyfert 1s \citep[e.g.][]{2005ApJ...630..122G,2006MNRAS.372..246P,2001AJ....122..549V}. In other works such as \cite{2015ApJS..217...26B}, they modelled the non-stellar continuum as a power-law with free parameters, but they used a single 11 Gyr, solar metallicity, single-burst spectrum from \cite{2003MNRAS.344.1000B} to model the host galaxy contribution.
For our sample we set the stellar and non-stellar contributions free in the code without prior over the steepness, the fraction of power$-$law, nor the stellar populations.
The results are shown in Figure \ref{Fig:Betas}. We find that the values of $\beta$ ranges between $-$2.9 and 0 with a mean value of $\beta = -1.6 \pm 0.6$ and a median of $-$1.6 with an IQR = 0.9. While for the contributions of the non-stellar component (right panel), we find a mean value of $F_{PL} = 0.61 \pm 0.22$ and a median of 0.64 with IQR = 0.29, in concordance with \cite{2006ApJS..166..128Z}. Contrary to what we expected, we do not find a correlation between $\beta$ and F$_{PL}$ with a Spearman rank order correlation coefficient of $r_s$ = $-$0.14 and a Student's t-test $T_s$ = $-$1.64 with a p-value = 0.24.

We checked our method by applying it to a sample of galaxies with low contribution from the AGN (as in the case of Seyfert 2) and to another sample with a high AGN contribution to the observed continuum (QSO). We constructed the sample of Seyfert 2 galaxies by taking 200 galaxies from the \cite{2010A&A...518A..10V} catalogued as Seyfert 2 randomly selected in the redshift range 0.01 < z < 0.1. For the QSO sample, we selected 100 QSO that are defined as isolated in the redshift range 0.2 < z < 0.31 from \cite{Donoso2018}. For the sample of Seyfert 2 galaxies, we obtained a F$_{PL}$ with a median of 0.19 and an IQR of 0.06.  This is in agreement with previously found results  \citep{1998MNRAS.297..579C,1999MNRAS.303..173S,2004MNRAS.355..273C}. In the case of the QSO sample, the median measured value of F$_{PL}$ was of 0.92 with an IQR of 0.30, which is in concordance with previous results \citep{2006AJ....131...84V,2016ApJ...816...85L,2017MNRAS.471...59L}. The results obtained for the F$_{PL}$ of these comparison samples are shown in Figure \ref{Fig:Betas}, right panel.

\begin{figure*}
\begin{minipage}{\linewidth}
\begin{center}
\includegraphics[trim = 0mm 0mm 0mm 55mm, clip, height=0.35\textheight]{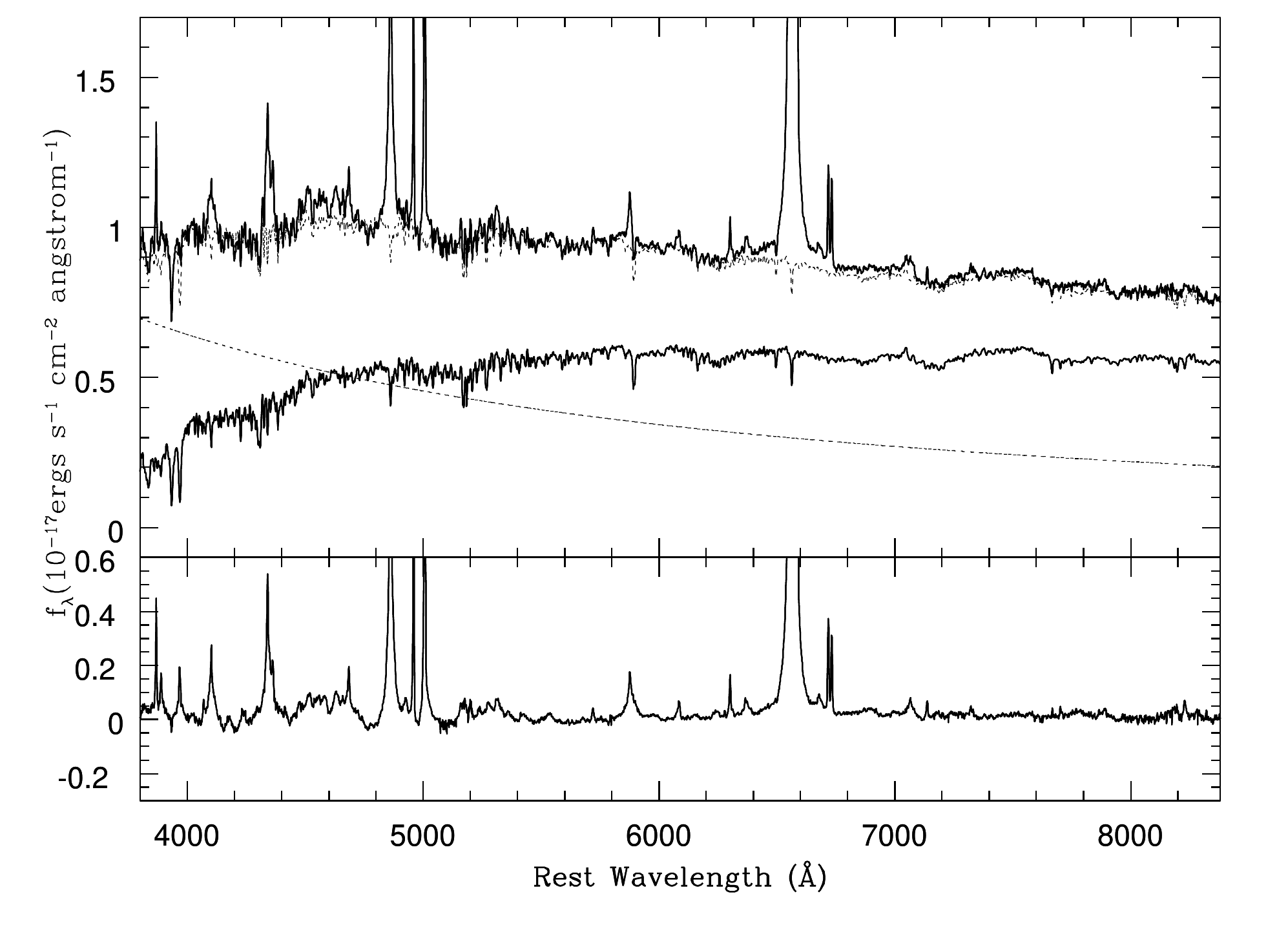}
\includegraphics[trim = 0mm 0mm 0mm 55mm, clip, height=0.35\textheight]{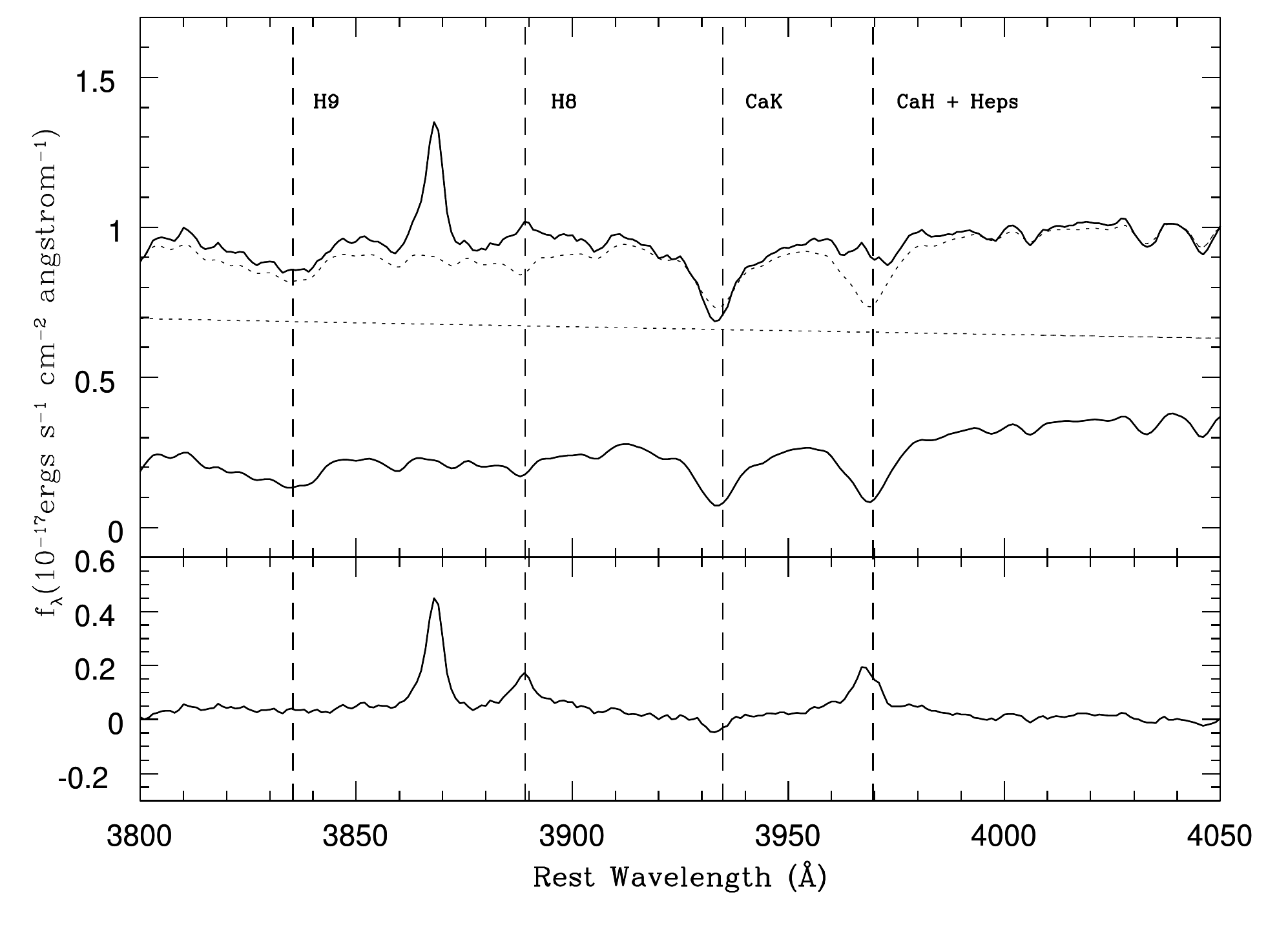}
\end{center}
\caption{(Top panel) Stacked spectra with geometric mean, (bottom panel) arithmetic mean. As in Figure \ref{Fig:fits2}, we plot the observed spectrum (thick solid line), the modelled spectrum (dotted line), the host galaxy spectrum (thin solid line), the power-law contribution (dotted line), and the residual spectra (lower panel). The observed spectra is normalised at the flux value in $\lambda$ = 4020. The bottom panel is the same as top panel, but is zoomed in on the blue part of the spectra. }
\label{fig:stacking}
\end{minipage}
\end{figure*}

Some spectral properties can be more evident in combined spectra of objects of the same class, as was pointed out by \cite{2001AJ....122..549V}. Combining spectra can also be an advantageous  tool that is often used to increase the S/N. Such composites have been studied on several occasions for quasars and BLAGNs \citep{2001PASA...18..221F,2005ApJ...633..638W,2017MNRAS.465...95P} in order to study the shape and variability of the continuum. We performed a stacking of the 131 spectra by using the geometrical mean value of the flux. 
As mentioned in \cite{2001AJ....122..549V}, the statistical method used to combine the spectra preserves different quantities. In this context, the geometric mean preserves the shape of the continuum. We performed a spectral synthesis on the geometric stacked spectra, and through equation \ref{eq:FPL}, we obtained a mean value of   $\beta_{(geo)}=-1.55$ with $F_{PL \ (geo)}$ = 0.63. The stacked spectra and model can be seen in Figure \ref{fig:stacking}.\\

\subsection{Emission lines}
\label{Broad Emission Lines}
Besides their characteristic continua, NLS1s exhibit some emission features that deserve attention. These include strong permitted Fe II emission lines in their optical and ultraviolet spectra, asymmetric [OIII] lines, and multiple component H${\beta}$ profiles \citep[e.g.][]{2008MNRAS.385...53M,2018A&A...615A..13S}.
The relative strength of the FeII multiplets is conventionally expressed as the following flux ratio of FeII to H$\beta$: $R_{4570}$ $\equiv$ Fe II $\lambda\lambda$ 4434 $-$ 4684 / H$\beta$, where FeII$\lambda\lambda$4434 $-$ 4684 denotes the flux of the FeII multiplets integrated over the wavelength range of 4434 $-$ 4684\AA{}, and H$\beta$ denotes the total flux of H$\beta$ (e.g. \citealt{2012AJ....143...83X, 2016MNRAS.462.1256C}).
We find that $R_{4570}$ ranges between 0.15 $-$ 2 and has a median of 0.74 with an IQR of 0.42. This value is higher than $R_{4570}$ $\sim$ 0.49 as found by \cite{2016MNRAS.462.1256C}. Nonetheless, our measurements are consistent with those of \cite{2006ApJS..166..128Z} and \cite{2007ApJ...670...60X,2012AJ....143...83X}, who found average values of 0.82, 0.75, and 0.7, respectively.

We explored whether there is a connection between the iron emission and the shape of the non-stellar continuum ($\beta$). We find no correlation between the F$_{PL}$ and R$_{4570}$ with a Spearman rank order correlation coefficient of $r_s$ = $-$0.15, Student's t-test $T_s$ = $-$1.51, and a p-value = 0.095. A mild correlation is found between the shape of the non-stellar continuum ($\beta$) and R$_{4570}$ with $r_s$ = 0.37, $T_s$ = 3.91, and a p-value = 0.0003. This would contribute to the hypothesis that photoionisation is not the only mechanism involved in the FeII emission \citep{2000ApJS..126...63R,2016MNRAS.462.1256C}.

\begin{figure*}
\begin{center}
\includegraphics[trim = 0mm 100mm 0mm 0mm, clip,width=\textwidth]{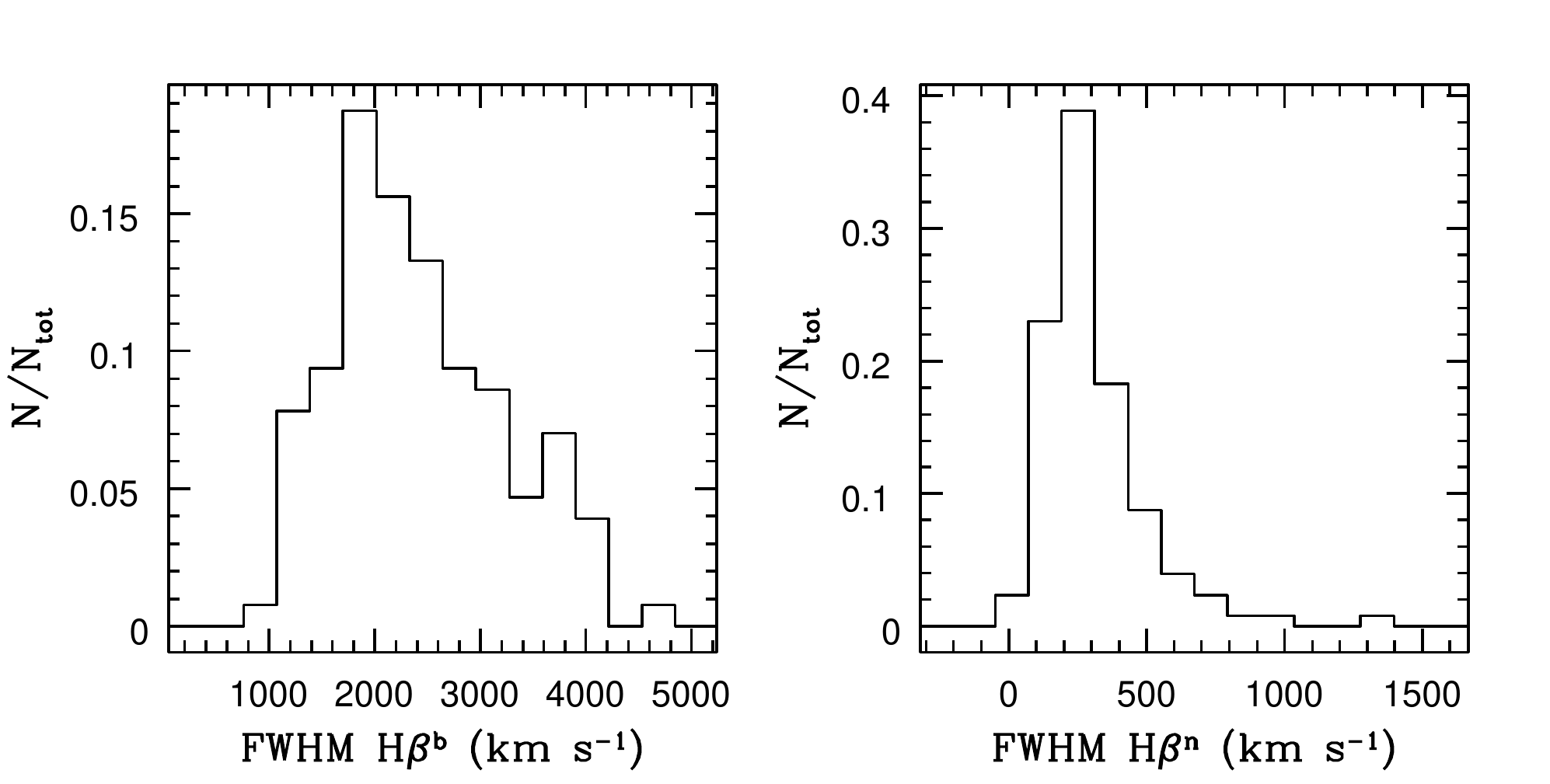}
\end{center}
\caption{FWHM distribution for broad (left) and narrow (left) component of H$\beta$. Units are in km s$^{-1}$ and were corrected by instrumental broadening.}
\label{fig:FW_Hb_OIII}
\end{figure*}

In Figure \ref{fig:FW_Hb_OIII}, we show the FWHM distribution for the broad and narrow component of H$\beta$. The left panel shows the broad component of H$\beta$ with values ranging between $\sim$ 900 $-$ 4500 km s$^{-1}$ , a median value of 2280 km s$^{-1}$ , and IQR = 1857 km s$^{-1}$.
The differences in FWHM from those measured by \citet{2006ApJS..166..128Z} seem to be mainly due to the different criteria adopted to fit the broad component of H$\beta$. However, we chose not to discard any of the galaxies with FWHM $>$ 2000 km s$^{-1}$ considering that AGNs with FWHM H$\beta$ $\leq$ 4000 km s$^{-1}$ belong to the same population \citep{2000ARA&A..38..521S,2018rnls.confE...2M}. 
Most galaxies present a narrow component FWHM from 0 to $\sim$800 km s$^{-1}$ with a median of 257 km s$^{-1}$ and an IQR = 174 km s$^{-1}$. As previously mentioned, two galaxies lack H$\beta$ in emission and we also found three 'outliers' with values above 800 km s$^{-1}$. They are 2MASX J22545221+0046316, SBS 0933+511, and 2MASX J08183571+2850224 with an FWHM of H$\beta^{n}$ of 890, 950, and 1390 km s$^{-1}$ , respectively.

\begin{figure*}
\begin{minipage}{\linewidth}
\begin{center}
\includegraphics[trim = 0mm 0mm 0mm 100mm, clip,width=\columnwidth]{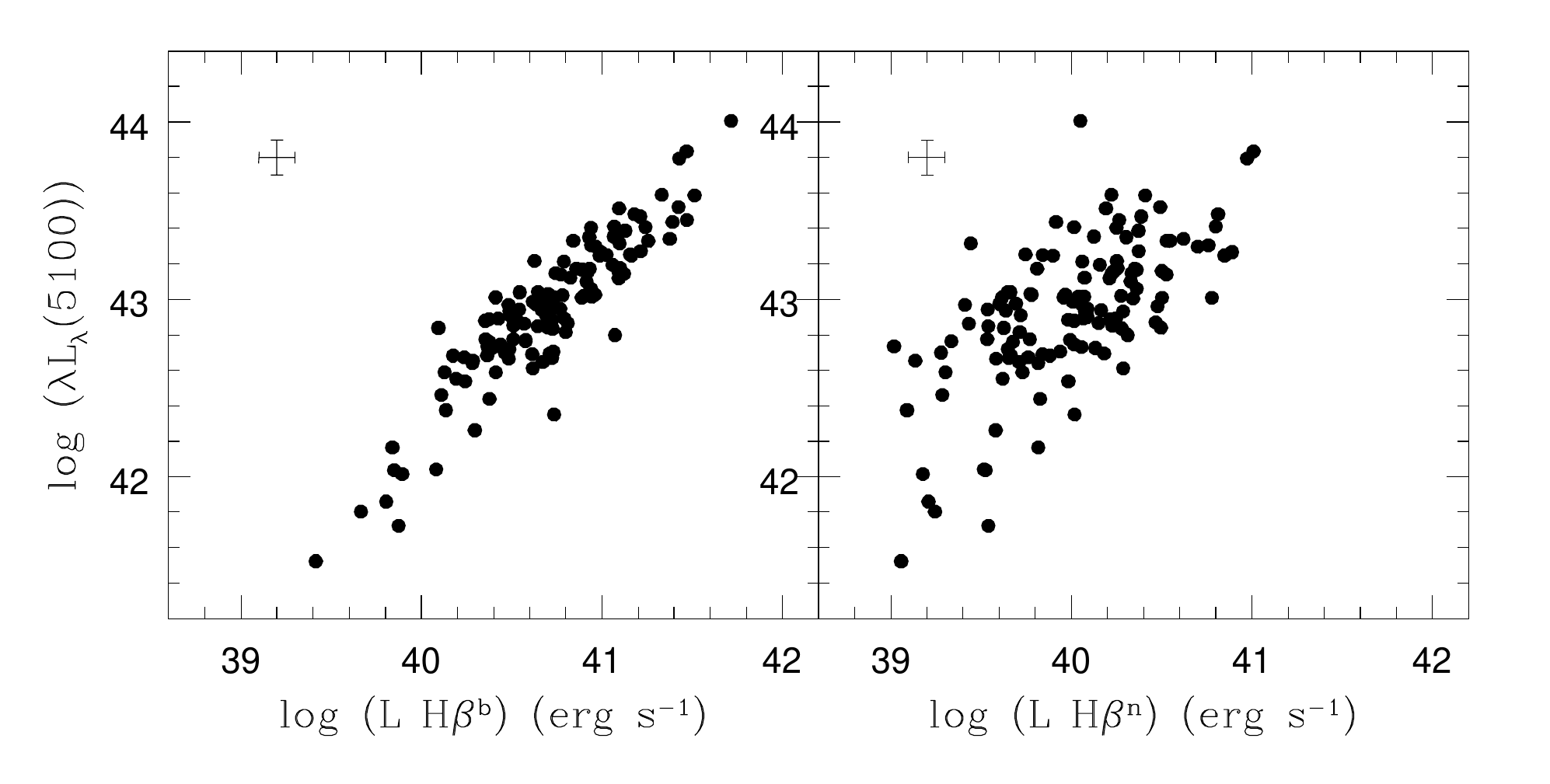}
\end{center}
\caption{Luminosity of featureless continuum in $\lambda$=5100 against the luminosity of broad (left) and narrow (right) component of H$\beta$.}
\label{LHb.fracpl}
\end{minipage}
\end{figure*}

We derived the luminosity $\lambda L_{\lambda}(5100)$ from the flux of the featureless continuum given by the power-law component and studied the relation between the strength of the AGN ($\lambda L_{\lambda}(5100)$) and the luminosity of the Balmer emission line H$\beta$. As expected for sources with emission that is dominated by photoionisation, a strong correlation was found between the luminosity of the broad component of H$\beta$ and $\lambda L_{\lambda}(5100)$ with a Spearman rank order correlation coefficient of $r_s$ = 0.88, a Student's t-test $T_s$ = 20.5, and p=3.4e-48.\ This result is in excellent agreement with \cite{2006ApJS..166..128Z,2016MNRAS.462.1256C}.
Considering the narrow component of H$\beta$, it follows the same trend but with a smaller and yet significant correlation coefficient of $r_s$ = 0.64, $T_s$ = 9.34, and p=2.6e-17 (right panel of Figure \ref{LHb.fracpl}). 

\subsection{Black hole masses and nuclear properties}
\label{Black Hole Masses and nuclear properties}

Supermassive black holes are of crucial interest in the study of AGNs. Determining their properties leads to the understanding of many mechanisms involved in the innermost regions of active galaxies. Several lines of evidence hint at small black hole masses ($M_{BH}$) as well as high accretion rates in NLS1 galaxies \citep[e.g.][]{2001NewA....6..321M,2009MNRAS.394.2141N,2015ApJ...801...38W}. According to \cite{2005ApJ...630..122G}, $M_{BH}$ can be estimated through luminosity and FWHM of the broad component of H$\beta$, as follows:

\begin{equation}
 M_{BH} = 3.26\times 10^6\left(\frac{L_{H\beta}}{10^{42}ergs^{-1}}\right)^{0.56}\left(\frac{FWHM_{H\beta}}{10^3kms^{-1}}\right)^2M_{\odot.}
 \label{eq:Greene}
 \end{equation}

This relation is very useful as it involves the luminosity and FWHM of the same Balmer emission line that are easily detectable even in distant AGNs. We determined $M_{BH}$ of the sample by using equation \ref{eq:Greene}, which takes the broad component of H$\beta$ into account, and after correcting for instrumental resolution. Since the errors in the measurements of the luminosity and FWHM are typically of 15\% and 10\%,  respectively, the error propagation for  $M_{BH}$ gives an uncertainty of $\sim$ 0.1 dex. 
In our sample, almost all of the galaxies have $M_{BH}$ spanning the values between $log(M_{BH}/M_{\odot})$= 5.6 $-$ 7.5 with a mean value in $log (M_{BH}/M_{\odot})$ = 6.5 $\pm$ 0.4 and a median value of 6.5 with IQR = 0.6. This is in agreement with previous results \citep[e.g.][]{2004ApJ...606L..41G,2006ApJS..166..128Z,2007ApJ...667L..33K,Schmidt2016,2016MNRAS.462.1256C}.

It has been demonstrated that $M_{BH}$ is tightly correlated with the stellar velocity dispersion of the bulge of normal galaxies, $\sigma_{\star}$ \citep{2000ApJ...539L...9F, 2000ApJ...539L..13G}. \cite{2004ApJ...615..652N} measured the bulge stellar velocity dispersion in 14 Seyfert 1 galaxies whose $M_{BH}$ were determined using the reverberation mapping technique and showed that the Seyfert galaxies followed the same $M_{BH} - \sigma_{\star}$ relation as non-active galaxies. Different studies of NLS1 galaxies have conflicting results about their location on the $M_{BH}$ - $\sigma_{\star}$ relation.
One of the main objectives of this work is to re-examine this relation using different methods to estimate $\sigma_{\star}$. It is important to note that for AGNs the stellar lines are usually diluted by the strong non-stellar continuum, and NLS1 are not an exception to that. Nonetheless, some absorption spectral features are evident in their spectra since we can detect stellar features like CaH+K, Mg, and CaT lines after careful inspection of the data (see sec. \ref{Sample}). We obtained an estimation of $\sigma_{\star}$ directly through a spectral synthesis technique by means of the code {\sc starlight}. This code performs a fit over all the stellar features and provides a $V_d$ parameter, which is actually a mean broadening parameter applied to the model that best fits the data. To get the proper  $\sigma_{\star,}$ we needed to correct this value from the instrumental and base spectral resolution as: $\sigma^2_{\star}=V_d^2-\sigma^2_{inst}+\sigma^2_{base}$.
We adopted the spectral resolution given by SDSS of $\sigma_{inst}$ $\sim$ 68km s$^{-1}$ in the vicinity of $\lambda=$ 5000 \AA{} and for the base spectra \citep[extracted from][]{2003MNRAS.344.1000B} $\sigma_{base} \sim$ 76 km s$^{-1}$ at the same wavelength.

Even though the region 7500\AA{} $-$ 9000\AA{} is affected by sky contamination, SDSS spectra residuals are similar in this range and in the Ca H+K vicinity. Additionally, the AGN contribution is not as powerful in the red region of the spectra, so the CaT lines are less diluted than the absorption lines in the blue region.  
In this scenario, \cite{2006ApJ...641..117G} argue that Ca triplet lines (CaT in 8498\AA{}, 8542\AA,{} and 8662\AA{}) provide the most reliable measurements of $\sigma_{\star}$ in AGNs. Furthermore, the width of the [OIII]$\lambda5007$ emission line is often used instead of stellar velocity dispersion due to observational difficulty of the latter. As stressed by \cite{2007ApJ...667L..33K}, the width of the [OIII]$\lambda$5007 line is a good surrogate for the stellar velocity dispersion when only the core of the line is considered.\ The same situation is true for NIR [SIII]$\lambda$9069 (\citealt{2009MNRAS.393..846V}).

\begin{figure*}
\begin{minipage}{\linewidth}
 \begin{center}
 \includegraphics[trim = 0mm 0mm 0mm 100mm, clip,width=\columnwidth]{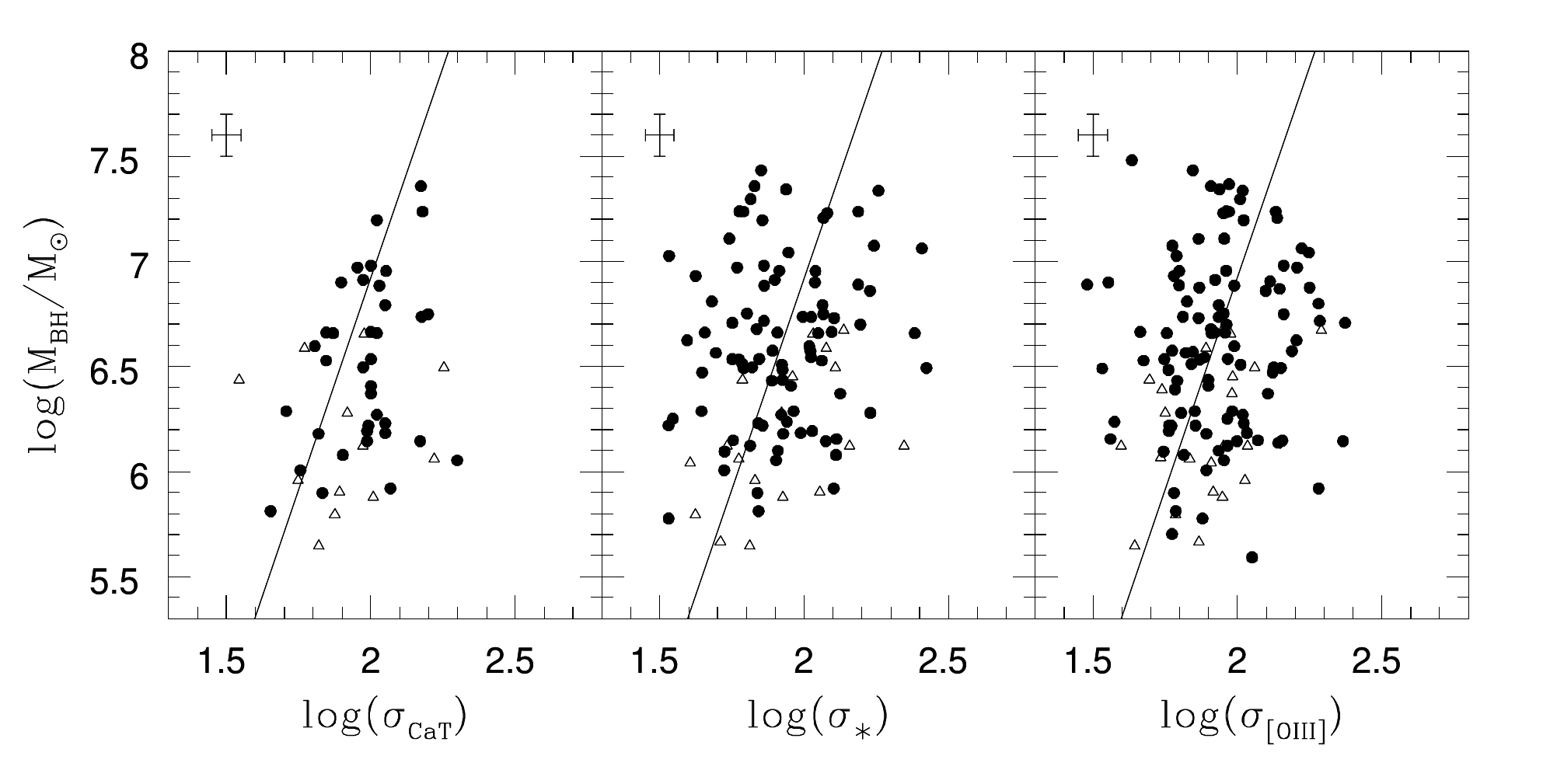}
 \end{center}
 \caption{Relation between black hole mass and stellar velocity dispersion measured from Calcium triplet ($\sigma_{CaT}$), $\sigma_{\star}$ obtained from {\sc starlight} and $\sigma_{[OIII]}$ from core component of [OIII]$\lambda$5007. Open triangles represent galaxies with L$_{bol}/L_{Edd}$ $>$ 1} and the solid line represents the relation given by Tremaine et al. (2002) for normal galaxies.
 \label{Fig:MBH_sigma}
\end{minipage} 
\end{figure*}

In Figure \ref{Fig:MBH_sigma} the locus of NLS1 in the $M_{BH} - \sigma_{\star}$ relation is seen with $\sigma_{\star,}$  which is obtained from the different approaches, as previously mentioned. The well-known $M_{BH} - \sigma_{\star}$ relation for normal galaxies, as parametrised by \cite{2002ApJ...574..740T},
is
\begin{equation}
 log\left(\frac{M_{BH}}{M_{\odot}}\right) = (8.13 \pm 0.06) + (4.02 \pm 0.32)log\left(\frac{\sigma_{\star}}{200km s^{-1}}\right)
 \end{equation}

and  is marked in the plots by the solid line. In the case of velocity dispersion, which is estimated from CaT lines (left panel in Figure \ref{Fig:MBH_sigma}), 67\% of the galaxies lie below the Tremaine line. By taking the uncertainties of our determinations of black hole masses and velocity dispersion into account, we obtained a more even distribution around the parametric relation when considering the velocity dispersion derived from {\sc starlight} and from the core component of [OIII]$\lambda$5007 (central and right panel of Figure \ref{Fig:MBH_sigma}). For $\sigma_{\star}$, 58\% of the objects lie below the Tremaine line, while in the case of the [OIII] line, NLS1s are closer to the relation and 46\% of the targets reside above it. This last result is in agreement with previous works \cite[e.g.][]{2001A&A...377...52W,2007ApJ...667L..33K,2016MNRAS.462.1256C}. Regardless of their spread in proximity from the Tremaine relation, these objects appear to be systematically below it with the higher deviation, as seen in the CaT lines, and we found no correlation between $M_{BH}$ and $\sigma$. This result is in agreement with the hypothesis that NLS1 reside in galaxies with pseudobulges \citep{2012ApJ...754..146M}.

\begin{figure}
\includegraphics[width=\columnwidth]{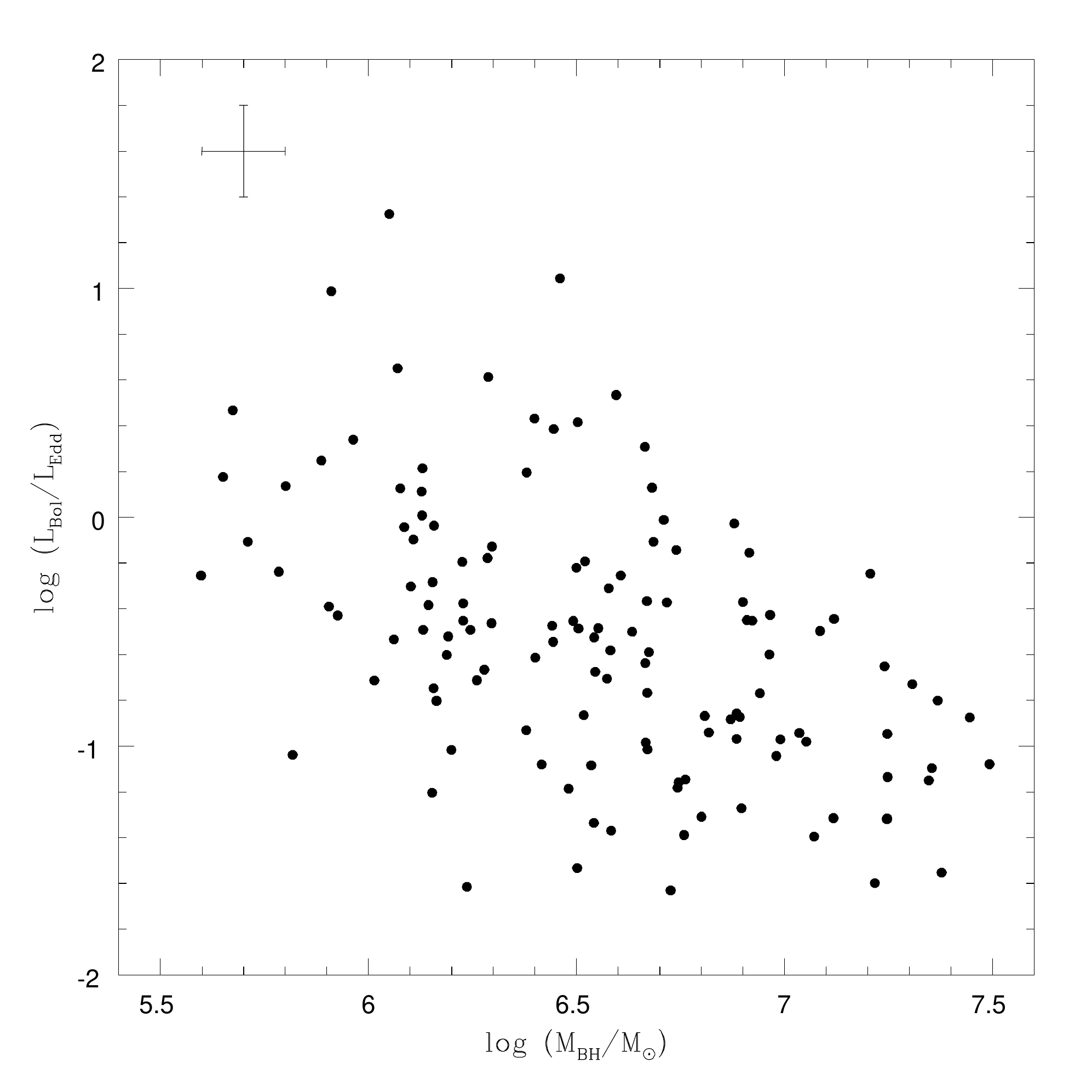}
\caption{Accretion rate relative to Eddington luminosity versus black hole mass}
\label{Fig:Acc_BHM}
\end{figure}

Black hole masses are closely linked to the accretion rate as is relative to the Eddington rate, which is generally parametrised as the ratio  $L_{bol}/L_{Edd}$ where the bolometric luminosity correction is assumed as $L_{bol} \sim 3500 \times L_{[OIII]}$ \citep{2004ApJ...613..109H} and $L_{Edd}=1.26 \times 10^{38} M_{BH}/M_{\odot}ergs^{-1}$. We find that several objects in our sample accrete at rates that are higher than, or close to,  the Eddington luminosity. Also an anti-correlation between the accretion rate and the black hole mass is observed ($r_s$ = $-$0.51, $T_s$ = $-$6.67, and a p-value = 1.1x10$^{-10}$), indicating that lower mass black holes grow faster than more massive ones, as shown in Figure \ref{Fig:Acc_BHM}. The high accretion objects seem to generally be located below the $M_{BH} - \sigma$ relation. In an evolutionary model, as proposed by \cite{2000MNRAS.314L..17M}, NLS1 black hole masses would move upwards on the $M_{BH} - \sigma$ plane.

\begin{figure}
\includegraphics[width=\columnwidth]{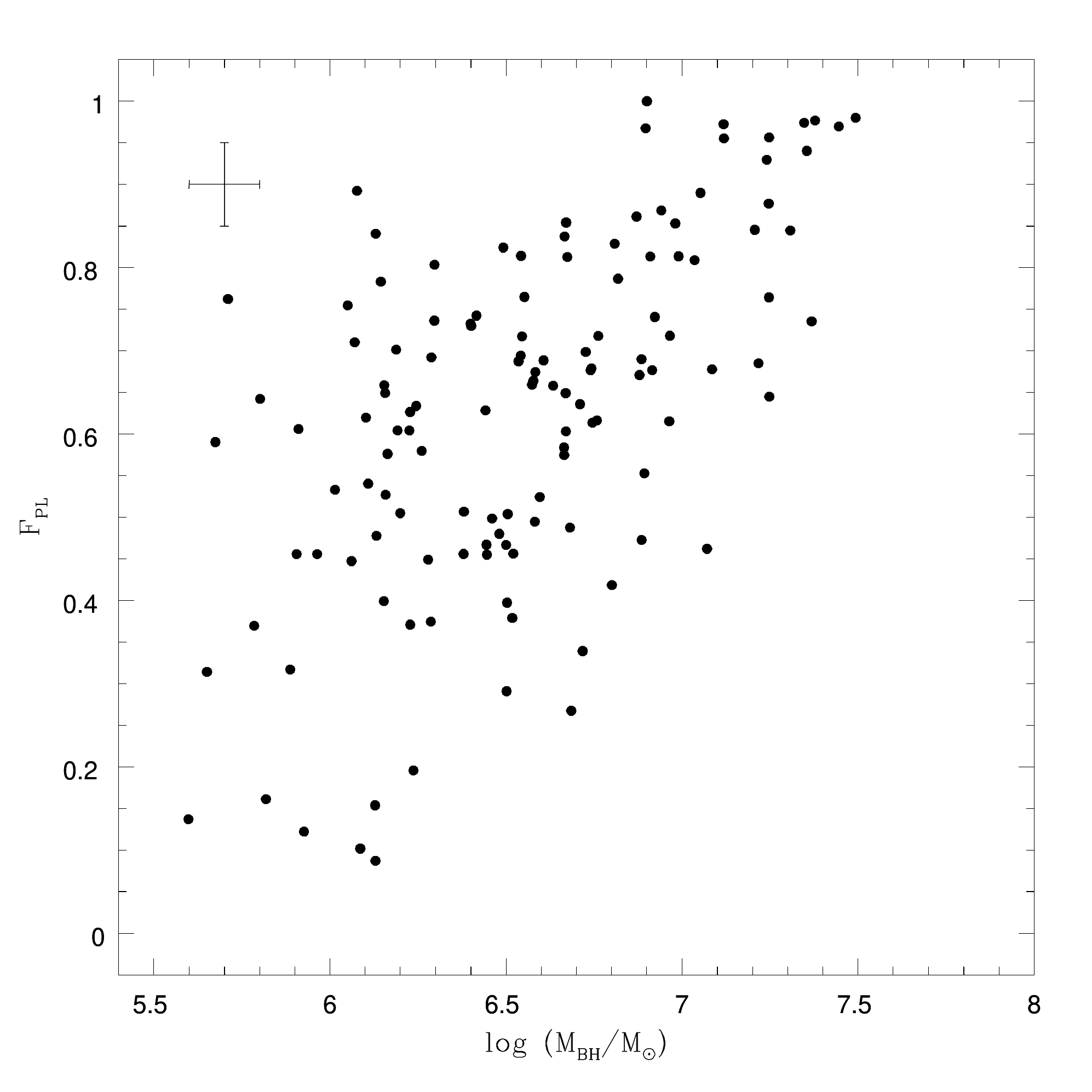}  
\caption{Fraction of power-law contribution versus black hole mass.}
 \label{MBH.fracpl}
\end{figure}

We compared the BH masses of our sample and the F$_{PL}$ of their total emission and we found the following interesting correlation between them: a Spearman rank order correlation coefficient of $r_s$ = 0.57, a Student's t-Test $T_s$ = 7.76, and a p-value = 1.7x10$^{-13}$. In Figure \ref{MBH.fracpl} it can be seen that, in general, objects with lower BH masses have lower F$_{PL}$ while galaxies with  F$_{PL}$ > 50 \% have BH masses $log(M_{BH}/M_\odot)$ > 7 in the high end of the relation.
The relation between the continuum luminosity and the central black hole mass is well established \citep{2000ApJ...533..631K,2002MNRAS.331..795M,2002ApJ...571..733V} with higher luminosities at greater BH mass. From our results, we see that the host galaxy contribution to the observed total spectra decreases at higher BH masses, that is, the continuum luminosity dilutes the stellar features of the spectrum.
 
\section{Discussion}
\label{Discussion}

The shape and strength of the continuum in NLS1 spectra is still a matter that is open for debate. Typically, the AGN continuum spectra are modelled as power-laws, and the slope of the continuum is assumed to be the same for all classes of AGNs. 
We found a large degeneracy for $\beta$ when the contribution of the non-stellar continuum is less than $\sim$ 40\%. Errors in the determination of the slope increase significantly when the power-law contribution is minor. Studying a sample of seven NLS1 galaxies using optical and near-IR CCD spectroscopy \cite{2000ApJS..126...63R} found an average value of $\beta$ = $-$1.41, while \cite{2016MNRAS.462.1256C} found a mean value of $\beta$ = $-$1.22 for a sample of 296 NLS1. Several works have computed the value for the spectral slope in AGN in the optical range from composite spectrum, such as \cite{1996PASA...13..212F,2001AJ....122..549V} who found mean values for $\beta$ of $-$1.65 and $-$1.55, respectively, from samples of QSO. \cite{2017MNRAS.465...95P} found a value of $-$1.89 from Seyfert 1 composite spectra, and they argued that the difference in the steepness of the spectra in wavelengths redder than 4000\AA \ is due to an increase in the contribution to the flux from the host galaxy. When we consider the contribution to the spectra by the featureless continuum F$_{PL,}$ it is evident that the host galaxy has a non-negligible impact on the observed spectra and that NLS1 galaxies are an intermediate case between QSO and Seyfert 2 galaxies (see Figure \ref{Fig:Betas}).

The well know relations between the central black hole and their host galaxy play a key role in understanding the black hole formation and evolution. Multiple pieces of evidence point towards a correlation between the M$_{BH}$ and the mass of the bulge \citep[e.g.][]{1995ARA&A..33..581K,1998AJ....115.2285M} and also a correlation with the stellar velocity dispersion \citep[e.g.][]{2000ApJ...539L...9F,2013ARA&A..51..511K}, suggesting that the velocity dispersion is an important parameter in understanding M$_{BH}$ evolution. As mentioned in sec. \ref{Intro}, there is still debate as to whether NLS1 follow the M$_{BH}$ $-$ $\sigma$ relation found for normal galaxies or not \citep[e.g.][]{2002ApJ...574..740T}. Some authors find that NLS1 galaxies do not follow this relation with the majority of the galaxies lying below the M$_{BH}$ $-$ $\sigma$ relation and with smaller M$_{BH}$ for a given stellar velocity dispersion \citep[e.g.][]{2001NewA....6..321M,2004ApJ...606L..41G,2005A&A...432..463M,2006ApJS..166..128Z,Schmidt2016}. In this scenario, it has been proposed that NLS1 are hosted with galaxies with pseudo-bulges, which is intrinsically different from normal galaxies \citep{2012ApJ...754..146M}.
Our results are in agreement with this, specially when considering the stellar velocity dispersion measured from CaT lines. Given the high accretion presented by the central black hole, NLS1 may be in an early evolutionary phase leading the black hole mass upwards in the M$_{BH}$ $-$ $\sigma$ plane and closer to the Tremaine relation.
Regarding Figures \ref{Fig:Acc_BHM} and \ref{MBH.fracpl}, accretion rates are usually found to be close to, or even higher than, the Eddington limit in NLS1 galaxies. But it has to be taken into account that the Eddington limit is proportional to the black hole mass. Furthermore, the Eddington limit is much lower in NLS1s than in Seyfert 1 or QSOs. Although NLS1s have high accretion rates (L$_{bol}$/L$_{Edd}$), the total luminosity in these objects is lower than in AGNs that host more massive black holes. In this way, the emission of the non-stellar continuum (FPL) is lower in NLS1 with low black hole masses, despite their high accretion. 

\section{Summary and final remarks}
In this paper we analysed the spectra of a sample of 131 NLS1 galaxies in the local Universe (z$<$0.1) taken from SDSS DR7 with the spectral synthesis technique. It allowed us to infer properties about the shape and strength of the non-stellar continuum and their relationship with other features of the spectra, such as the emission lines. The main results of this paper are summarised as follows:\\
Firstly, for the galaxies in our sample, we obtain a median spectral index of $\beta$ = $-$1.6 and a median fraction of non-stellar continuum of 0.64. We applied our method to two different samples, one consisting of objects with an expected high AGN contribution to the continuum (QSO), and another one with low AGN contribution to the continuum (Seyfert 2). In the case of QSOs, we obtain a median F$_{PL}$ = 0.92, while for Seyfert 2 F$_{PL}$ = 0.19.
Secondly, NLS1 galaxies are known to have strong FeII emission, we find a range of R$_{4570}$ between 0.15 and 2 with a median of 0.74. We find no correlation among R$_{4570}$ and F$_{PL}$ and only a weak correlation with $\beta$.
Thirdly, in concordance with previous studies, the galaxies in our sample occupy the M$_{BH}$ $-$ $\sigma$ plane falling below Tremaine's relation. This separation is more evident when the stellar velocity dispersion is measured from Calcium II Triplet lines.
Finally, we find a new correlation between the black hole mass and the fraction of non-stellar contribution to the continuum. Related to this, galaxies with higher black hole masses tend to present a higher amount of non-stellar emission.

In the context of the unified model, these kind of galaxies present intermediate F$_{PL}$, lower M$_{BH,}$ and also show lower stellar velocity dispersion.\ This implies that NLS1 are an extension of BLAGNs at the lower mass end.

\section*{Acknowledgements}

G.A.O. and E.O.S. want to thank Damian Mast for fruitful discussions. We also appreciate the helpful comments and suggestions made by the anonymous referee, which improved this article. This work was partially supported by Consejo de Investigaciones Cient\'ificas y T\'ecnicas (CONICET) and Secretar\'ia de Ciencia y T\'ecnica de la Universidad Nacional de C\'ordoba (SecyT).
Funding for the SDSS and SDSS-II has been provided by the Alfred P. Sloan Foundation, the Participating Institutions, the National Science Foundation, the U.S. Department of Energy, the National Aeronautics and Space Administration, the Japanese Monbukagakusho, the Max Planck Society, and the Higher Education Funding Council for England. The SDSS Web Site is \url{http://www.sdss.org/.}
The SDSS is managed by the Astrophysical Research Consortium for the Participating Institutions. The Participating Institutions are the American Museum of Natural History, Astrophysical Institute Potsdam, University of Basel, University of Cambridge, Case Western Reserve University, University of Chicago, Drexel University, Fermilab, the Institute for Advanced Study, the Japan Participation Group, Johns Hopkins University, the Joint Institute for Nuclear Astrophysics, the Kavli Institute for Particle Astrophysics and Cosmology, the Korean Scientist Group, the Chinese Academy of Sciences (LAMOST), Los Alamos National Laboratory, the Max-Planck-Institute for Astronomy (MPIA), the Max-Planck-Institute for Astrophysics (MPA), New Mexico State University, Ohio State University, University of Pittsburgh, University of Portsmouth, Princeton University, the United States Naval Observatory, and the University of Washington.
The {\sc starlight} project is supported by the Brazilian agencies CNPq, CAPES and FAPESP and by the France-Brazil CAPES/Cofecub programme.

\bibliography{bibliography.bib}

\end{document}